# General Multi-State Rework Network and Reliability Algorithm

Zhifeng Hao
Schools of Mathematics and Big Data
Foshan University, Foshan, China
mazfhao@fosu.edu.cn

Wei-Chang Yeh*
Integration & Collaboration Laboratory
Department of Industrial Engineering and Management Engineering
National Tsing Hua University, Hsinchu, Taiwan
yeh@ieee.org

Zhenyao Liu
Integration & Collaboration Laboratory
Department of Industrial Engineering and Management Engineering
National Tsing Hua University, Hsinchu, Taiwan
liuzhenyao49@gmail.com

Abstract – A rework network is a common manufacturing system, in which flows (products) are processed in a sequence of workstations (nodes), which often results in defective products. To improve the productivity and utility of the system, the rework network allows some of the defective products to go back to the "as normal" condition after the rework process. In a recent study, Song proposed an algorithm to correct more than 21 archive publications regarding the rework network reliability problem, which is an important real-life problem. However, we prove that Song's proposed algorithm is still incorrect. Additionally, we provide an accurate general model based on the novel state distribution with a smaller number of limitations. Furthermore, we propose an algorithm to calculate the reliability of the multi-state rework networks using the proposed novel state distributions.

Keywords: Analytical results; Discrete-event simulation; System reliability; Rework

## 1. INTRODUCTION

The rework network is a special network with a rework process, which is popular in reworking defective products back to the "as normal" condition [1-7]. A traditional rework network can be considered to be a simple network but with both the deterioration effect [8-10] and the learning effect [11-13]. During the transmission and processing of the flow (products) from one node (such as workstation, machine, or worker) to another in the rework network, the amount of flow decreases owing to the deterioration effect of the discarded/defective products [8-10]. However, in some nodes, the amount of flow increases owing to the learning effect [11-13], resulting from some detective products being able to be processed again after the rework to fix/repair them.

A rework network $G(V, E, \mathbf{D})$ comprises a node set $V = \{1, 2, …, n\}$ and an arc set $E = \{e_{i,j} \mid i, j \in V\}$. $\mathbf{D}$ is the state distribution defining the state levels $\{0, 1, …, \mathbf{D}_{max}(k)\}$ and the occurrent probability $\mathbf{D}(k, l) = p_{k,l}$ of the related state level $l \in \{0, 1, …, \mathbf{D}_{max}(k)\}$ and the defect rate $\mathbf{D}(k) = p_k$ for all nodes $k \in V$. In $G(V, E, \mathbf{D})$, each node could be either a process station, work station, unit, machine, tool, or worker to process (including any type of manufacture) the products that are input to the rework network, and the flow could be any product, such as oil or a car [14-16].

Let node $\beta$ be one of the previous nodes of node $\alpha$, where $\alpha \neq \beta$ and $\alpha, \beta \in V$. For simplicity, nodes $\alpha$ and $\beta$ are referred to as the rework-input node and the rework-begin node to represent the node that sends defects to the rework process and the first node in the rework process, respectively. Hence, node $\alpha$ sends the defective products back to node $\beta$ to be reworked in the rework process.

Note that [1,2]:
1. Except for node $\alpha$, all the other nodes discard the defective products immediately after they are identified.
2. Each defect has only one chance to be reworked, i.e., it cannot be reworked again if it is fails after it has been sent back from node $\alpha$.

For example, the network illustrated in Figure 1 is a rework network $G(V, E, \mathbf{D})$, where $V = \{1, 2, 3, 4\}$ and $E = \{e_{1,2}, e_{2,3}, e_{3,4}\}$ [1, 2]. The state distribution $\mathbf{D}$ is listed in Table 1. Node 0, indicated by a dotted circle is a pseudo node (i.e., not real), which is used to express the values of input flows.

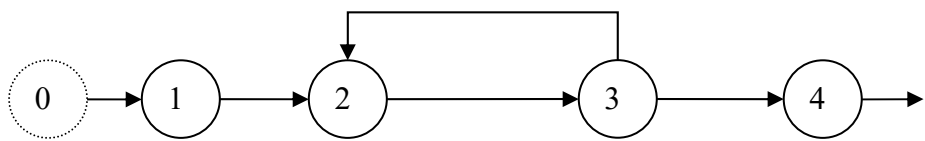

**Figure 1.** Example rework network.

**Table 1.** State distribution **D** of Figure 1.

| k | **D**(k)=$p_k$ | State l | **D**(k, l)=$p_{k,l}$ | k | **D**(k)=$p_k$ | State l | **D**(k, l)=$p_{k,l}$ |
|---|---|---|---|---|---|---|---|
| 1 | 0.9 | | | 3 | 0.9 | | |
| | | 0 | 0.01 | | | 0 | 0.005 |
| | | 5 | 0.01 | | | 3 | 0.005 |
| | | 10 | 0.05 | | | 6 | 0.010 |
| | | 15 | 0.93 | | | 9 | 0.010 |
| | | | | | | 12 | 0.010 |
| | | | | | | 15 | 0.960 |
| 2 | 0.9 | | | 4 | 0.9 | | |
| | | 0 | 0.01 | | | 0 | 0.005 |
| | | 5 | 0.02 | | | 3 | 0.005 |
| | | 10 | 0.02 | | | 6 | 0.005 |
| | | 15 | 0.95 | | | 9 | 0.010 |
| | | | | | | 12 | 0.015 |
| | | | | | | 15 | 0.960 |

Among these four nodes, there is one special arc from node 3 to node 2 in Fig. 1 to represent the rework process. Node 3 is the rework_input node α and node 2 is the rework_begin node β. Defects are only allowed to be sent back from node 3 to node 2 to be reworked, and the other nodes simply discard the defective products. For example, any defective product must be discarded if it is detected right after being processed in nodes 1, 2, or 4; however, it can be sent back to node 2 to be reworked if it is detected in node 3.

A path in a rework network is a procession to describe the way to process products in a rework network. For example, there are only two paths in Fig. 1: the normal path $p_n$ = {1, 2, 3, 4}, which represents the process of perfect products without any defect, and the rework path $p_r$ = {1, 2, 3, 2, 3, 4}, which represents the process of defective products by the rework process [1, 2].

In a recent study, Song pointed out that there exists an error in all the current publications related to the rework problem in network reliability, and proposed a rework network model, wherein the number of defective products generated in each node followed the binomial distribution [1, 2]. For example, in node $i$, the probability of having $p_i$ and $q_i$ units of perfect and defective products is $C_{q_i}^{p_i+q_i}(1-\delta_i)^{p_i}\delta_i^{q_i}$, where $\delta_i$ is the defect rate of node $i$. Song also proposed a method for computing the multi-state two-terminal reliability in such a network through rework operations [1].

It is claimed that the proposed algorithm has the following advantages: (1) The algorithm's accuracy has been verified by using a discrete-event simulation, and a mathematical analysis of the errors published in previous papers has been provided. (2) The algorithm is a useful tool for assessing the validity of any future proposed approach for other stochastic reliability problems.

There are indeed some errors published in the more than 25 previous related papers cited in Song's paper that have been corrected [1]. However, the concept of the binomial rework network is still ambiguous and the proposed algorithm is still incorrect due to fatal errors in Eqs.(14)–(17) of Page 262 in [1]. As there are several important applications of the rework network in real-life as demonstrated in [1, 2] and all related references provided in [1, 2], there is a need to provide a general rework network with a correct algorithm to calculate its reliability.

A more general rework network is proposed by removing some impractical restrictions from the traditional rework network together with a novel algorithm for solving the reliability of the proposed generalized binomial rework network. The remainder of this paper is structured as follows. Section 2 presents the required notations of the proposed general rework network. The details of the ambiguities in Song's rework network and the errors of its related algorithm are provided in Section 3. The proposed general rework network is discussed in Section 4 together with its assumption, innovations, and a novel algorithm for its reliability. Section 5 presents a performance analysis of the proposed algorithm. Finally, the conclusion is presented in Section 6.

## 2. NOTATIONS AND ASSUMPTIONS

The required notations for the proposed general rework network and reliability algorithm are described in this section.

$V$:    node set $V = \{1, 2, …, n\}$

$E$:    arc set $E = \{e_{i,j} \mid i, j \in V\}$

$\delta_k$:    defect rate of node $k \in V$ if not in the rework process

$\delta$:    rate of sending defects in the rework process

| | | |
|---|---|---|
| $\gamma_k$: | defect rate of node $k \in V$ in the rework process |
| $\mathbf{D}_{max}(k)$: | maximal state level of node $k \in V$. |
| $\mathbf{D}_k$: | state levels $\mathbf{D}_k = \{0, 1, \ldots, \mathbf{D}_{max}(k)\}$ |
| $\mathbf{D}(k, l)$: | occurrent probability of the related state level $l \in \mathbf{D}_k$. |
| $\mathbf{D}$: | state distribution defining $\mathbf{D}_{max}(k)$, $\mathbf{D}(k, l)$, $\mathbf{D}_k$, $\delta_k$, $\delta$ and $\gamma_k$ for all $k \in V$. |
| $c_i(l)$: | smallest state level of node $i$, which is not less than $l$, e.g., $c_2(2) = 5$ in Table 1. |
| $C_i(l)$: | probability of having $c_i(l)$, e.g., $C_2(2) = 0.02$ in Table 1. |
| $\alpha$: | rework-input node that sends the defective products to the rework process, e.g., node 3 in Fig. 1 |
| $\beta$: | rework-begin node, which is the first node in the rework process, e.g., node 2 in Fig. 1 |
| $G(V, E, \mathbf{D})$: | rework (activity-on-node) network with $V = \{1, 2, \ldots, n\}$, $E = \{e_{i,j} \mid i, j \in V\}$, $\mathbf{D}$, $\alpha \in V$, and $\beta \in V$ |
| $P(\bullet)$: | probability of the occurrence of the event $\bullet$ |
| $b$: | input value |
| $b^*$: | input value to the rework node |
| $d$: | output demand |
| $R_d$: | system reliability that at least $d$ units of output do not have any defect |
| $F_o$: | total output |
| $p_n$: | normal path, which is a path without rework |
| $p_r$: | rework path |
| $F_o(p)$: | output of the path $p$ |
| $F_o(p, k)$: | output of node $k$ for path $p$ |
| $F_i(p, k)$: | input of node $k$ for path $p$. Note that $F^*(p_n, k) = \min(F(p_n, j), \mathbf{D}_{max}(k))$ if node $j$ is the node right before node $k$ in $p_n$. |
| $p_x, p_{x,i}$: | non-rework vector $p_x = (p_{x,0}, p_{x,1}, p_{x,2}, \ldots, p_{x,n})$, which is the state vector without |

rework and $p_{x,i}$ is the number of perfect products in node $i$ without using the rework process for $i = 1, 2, \ldots, n$, where $p_{x,0} = b$ and $p_{x,i} \leq p_{x,j}$ if $j \leq i$.

$q_x$, $q_{x,i}$:    $q_y = (q_{x,0}, q_{x,1}, q_{x,2}, \ldots, q_{x,n})$ and $q_{x,i} = b - \sum_{k=1}^{i} p_{x,k}$ if $p_x$ is the related non-rework vector

$\pi_y$, $\pi_{y,j}$:    rework vector $\pi_y = (\pi_{y,\alpha}, \pi_{y,\beta}, \pi_{y,\beta+1}, \ldots, \pi_{y,n})$, which is the state vector right after rework and $\pi_{y,j}$ is the number of perfect products produced in node $j$ from the rework process for $j = \beta, \beta+1, \ldots, n$, where $\pi_{y,\alpha} = b^*$ and $\pi_{y,i} \leq \pi_{y,j}$ if $j \leq i$

$\theta_y$, $\theta_{y,j}$:    $\theta_y = (\theta_{y,\alpha}, \theta_{y,\beta}, \theta_{y,\beta+1}, \ldots, \theta_{y,n})$ and $\theta_{y,i} = b^* - \sum_{k=\beta}^{i} \pi_{y,k}$ if $\pi_y$ is the related rework vector

$C_q^Q$:    $\dfrac{Q!}{q!(Q-q)!}$

## 2. ERRORS AND UNCLEAR SITUATIONS IN PREVIOUS PAPERS

The current algorithms in calculating the rework network reliability are based on Eqs.(1)–(3) (in this paper) [1, 2]. Note that Eqs. (1)–(3) are the same as Eqs. (14)–(16) in [1, 2], and Eqs.(16) and (17) are duplicates in [1, 2]. However, this does not affect the final results.

$$R_d = P(F_o \geq d) \tag{1}$$

$$= P(F(p_n) \geq d) + \sum_{j=1}^{d} [P(F(p_n) = d - j) \cdot P(F(p_r) \geq j)] \tag{2}$$

$$\sum_{j=1}^{d} [P(F(p_n) = d - j) \cdot P(F(p_r) \geq j)] = \sum_{b_n = d-j}^{d-j} \sum_{a_n = b_n}^{b} \sum_{b_{n-1} = a_n}^{b} \sum_{a_{n-1} = b_{n-1}}^{b} \cdots \sum_{b_2 = a_3}^{b} \sum_{a_2 = b_2}^{b} \sum_{b_1 = a_2}^{b}$$

$$\sum_{a_1 = b_1}^{b} \left( \left[ \prod_{i=1}^{n} P(A_i = a_i \mid B_{i-1} = b_{i-1}) \cdot P(B_i = b_i \mid A_i = a_i) P(F(p_r) \geq j) \right] \right). \tag{3}$$

The details of the three errors and one faulty assumption (named Assumption 3 in [1]) in [1, 2] are discussed below.

## 2.1 First Error

Let $p_n$ and $p_r$ be a normal path and a rework path, respectively, and $F(\bullet)$ be the output of path $\bullet$. The following lemma proves that Eq.(2) (i.e., Eq.(15) in [1,2]) is incorrect.

**Lemma 1.** The equation listed in Eq.(2) which is the basis of Song's algorithm is incorrect.

**Proof.** The reliability of having at least $d$ units of output without any defect is $R_d = P(F_o \geq d)$. The total output $F_o$ includes the perfect products that do not require any rework ($F_o(p_n)$) and the defective products that became non-defective after the rework ($F_o(p_r)$). Hence,

$$R_d = P(F_o \geq d) = P(F_o(p_n) + F_o(p_r) \geq d). \tag{4}$$

Based on the inclusive-exclusive theorem [17-22], we have

$$P(F_o(p_n) + F_o(p_r) \geq d) = P(F_o(p_n) \geq d) + P(F_o(p_r) \geq d) - P(F_o(p_n) \geq d \text{ and } F_o(p_r) \geq d). \tag{5}$$

However, Eq. (5) is not equal to Eq.(2), i.e.,

$$R_d = P(F_o(p_n) + F_o(p_r) \geq d)$$

$$= P(F_o(p_n) \geq d) + P(F_o(p_r) \geq d) - P(F_o(p_n) \geq d \text{ and } F_o(p_r) \geq d)$$

$$\neq P(F_o(p_n) \geq d) + \sum_{j=1}^{d}[P(F_o(p_n) = d - j) \cdot P(F_o(p_r) \geq j)]. \tag{6}$$

Thus, Eq. (2) is incorrect. □

Thus, all of Songs' results, models, and algorithms are incorrect because they are all based on Eq.(2).

## 2.2 Second Error

In Eq.(3), i.e., Eq.(16) in [1,2], the probability of the outputs of $p_n$ and $p_r$ are treated to be independent, i.e.,

$$P(F_o(p_n) \geq d \text{ and } F_o(p_r) \geq d) = P(F_o(p_n) \geq d) \times P(F_o(p_r) \geq d). \tag{7}$$

However, Eq. (7) is incorrect due to either of the following reasons: (i) The capacity of each workstation is shared by both the normal path and the rework path and hence, the probability of the outputs of both the paths cannot be treated independently. (ii) We require more assumptions to determine how the normal and rework paths share the capacities of the nodes.

**2.3 Third Error**

The system (network) reliability is defined as the probability of the system being able to complete some predefined goals under pre-requested conditions. For example, the reliability of the rework network is defined in Eq.(4), which is the probability of the rework network to have at least $d$ successful outputs. However, in [1, 2], the reliability is misunderstood such that

$$R_d = \mathrm{P}(F_o(p_n)+F_o(p_r) =d). \tag{8}$$

**2.4 Unclear Situations**

Majority of the current algorithms are incorrect due to the faulty Assumption 3 listed in [1] as follows.

> "(3) The number of rework items reprocessed in $WS_i$ in the $v$th rework path depends on the remaining random capacity in $WS_i$ after the regular path and the first, second, …, ($v$-1)th rework paths are completed, where $i = k_v, k_v + 1, \ldots, n$."

Note that $WS_i$ is denoted as node $i$ in this study.

There are two cases that can be considered after the regular path is completed, and both these cases are inconsistent with Assumption 3 listed above.

Case 1.  The capacities of all the nodes in the regular path are released from these rework items in the regular path. In such a case, it is trivial that the number of rework items in node $i$ unnecessarily depends on the remaining capacity in node $i$, i.e., the results derived based on the above Assumption 3 are incorrect.

Case 2.  The capacities of the nodes in the regular path are still partly occupied by the rework items in the regular path. In this case, the number of new items entering the system depends on the remaining capacity in node $i$, resulting in the reliability reducing to zero from time to time because the capacity of each working station is reduced and occupied by old input items, unless there are no new items required to be input into the network, which is impractical. Hence, the results obtained in [1] are incorrect.

From the above three errors and the faulty Assumption 3 listed in [1], there is currently no existing exact-solution algorithm for such a rework network reliability problem.

## 3. PROPOSED GENERALIZED BINOMIAL REWORK NETWORK

The proposed rework network generalized from the traditional rework network is discussed in this section. We first address the limitation of the current rework network, and then generalize the rework network by considering more practical situations and providing more robust assumptions to fix some of the ambiguous results of the current rework network. Moreover, we provide solutions based on the multi-vector to denote the situations where the rework network is able to produce the required amount of products to help in developing algorithms and calculating the rework network reliability.

### 3.1 Innovation and Assumptions of the proposed rework network

To fix the unclear situations and errors arising in the current rework network, the following assumptions are provided:

**Assumption 1.** The rework process is unnecessary for the defect products if and only if the number of all perfect products is greater than or equal to $d$. In the current rework network, there is no clear answer whether a rework is still needed if the number of final non-defective products are already met.

**Assumption 2.** If the rework process is required, it is implemented only after the normal processes of all the products are completed. It is unclear as to how the normal products and reworked products share the capacities of each node, as discussed in Section 2.4 [1, 2].

**Assumption 3.** The defect rates of all the products during the rework process is greater than those of the products before the rework process, i.e., $\delta_i \leq \gamma_i$ for all $i \in V$. Note that $\delta_i = \gamma_i$ for all $i \in V$ in the current rework network [1, 2].

**Assumption 4.** Even in node α, some defects cannot be sent back to node β for a rework process and must be discarded, i.e., $\delta \leq 1$. Note that $\delta = 1$ in the current rework network, which is impractical [1, 2].

**3.2 Multi-vector structure for the solution**

A feasible solution is to identify a successful manner in which $b$ units of flow can be sent into node 1 through which at least $d$ units of flow are received from node $n$ [23-25]. Based on the proposed Assumption 1, a feasible solution must have a rework process only if the total number of perfect products outputted from node $n$ is less than $d$.

From the proposed Assumption 2, two consecutive processes are required for the rework feasible solution. Hence, two special vectors called non-rework vector and rework vector for the first and second processes are proposed rather than the normal and rework paths used in the current rework network, respectively. The number of rework and non-rework vectors are still growing exponentially with the network size.

Let $p_x = (p_{x,0}, p_{x,1}, p_{x,2}, \ldots, p_{x,n})$ and $q_x = (q_{x,0}, q_{x,1}, q_{x,2}, \ldots, q_{x,n})$ be two non-rework state vectors (without using the rework process), where $p_{x,i}$ and $q_{x,i}$ are the number of perfect and defect products right after processing in node $i$ for $i = 1, 2, \ldots, n$, respectively. For example, in Figs. 2 and 3(a), (14, 14, 12, 11, 10) and (14, 13, 12, 5, 5) are, respectively, two non-rework state vectors of the rework network shown in Fig. 1.

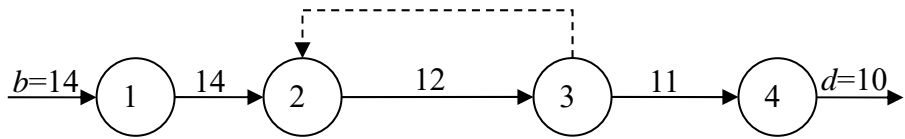

**Figure 2.** Feasible solution $X = (p_x, 0)$, where $p_x = (14, 14, 12, 11, 10)$.

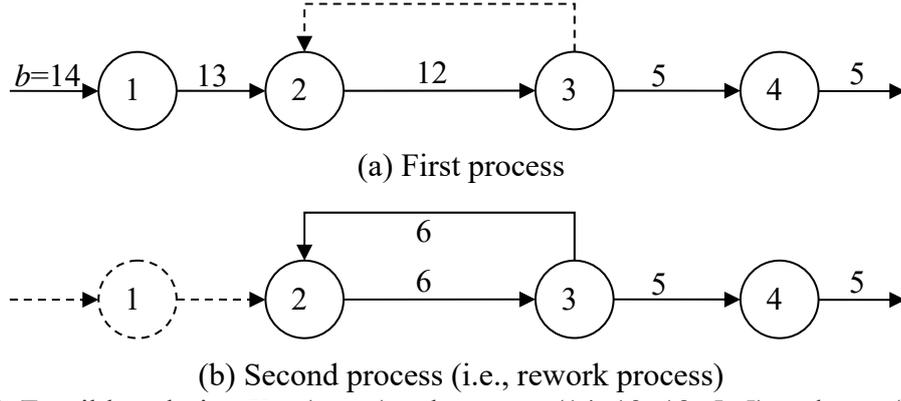

(a) First process

(b) Second process (i.e., rework process)

**Figure 3.** Feasible solution $X = (p_x, \pi_y)$, where $p_x = (14, 13, 12, 5, 5)$ and $\pi_y = (6, 6, 5, 5)$.

Similarly, let $\pi_y = (\pi_{y,\alpha}, \pi_{y,\beta}, \pi_{y,\beta+1}, \ldots, \pi_{y,n})$ and $\theta_y = (\theta_{y,\alpha}, \theta_{y,\beta}, \theta_{y,\beta+1}, \ldots, \theta_{y,n})$ be two rework state vectors, where $\pi_{y,j}$ and $\theta_{y,j}$ are the number of perfect and defect products right after processing in node $i$ for $i = \beta, \beta+1, \ldots, n$ during the rework process, respectively. Note that $p_{x,0} = b$, $\pi_{y,\alpha} = b^*$, and $q_{x,0} = \theta_{y,0} = 0$. For example, as shown in Fig. 3(b), $\pi_y = (6, 6, 5, 5)$ is a rework state vector of the rework network shown in Fig. 1 and six defects are sent back from node $\alpha = 3$ to node $\beta = 2$ for the rework process, six, five, and five perfect products are outputted from nodes 2, 3, and 4, respectively.

The following two lemmas discusses the relationships between $p_x$ and $q_x$.

**Lemma 1.** $p_{x,i} \leq p_{x,j}$ and $q_{x,j} \leq q_{x,i}$ if $j \leq i$.

**Proof.**  Based on the deterioration effect [8-10], the number of perfect products decreases with the number of processed nodes if no reworked process is conducted. Hence, we have $p_{x,i} \leq p_{x,j}$ if $j \leq i$. Similarly, the number of defects are increasing owing to the augmentation effect and we have $q_{x,j} \leq q_{x,i}$ if $j \leq i$.

**Lemma 2.**

$$p_{x,i} = b - \sum_{k=1}^{i} q_{x,k} = p_{x,i-1} - q_{x,i} \tag{9}$$

$$q_{x,i} = b - \sum_{k=1}^{i} p_{x,k} = p_{x,i-1} - p_{x,i}, \tag{0}$$

respectively.

**Proof.** The number of perfect products right after being processed in node 1 is the total number of products minus the number of defect products after being processed in node 1, i.e., $p_{x,1} = b - q_{x,1}$. Similarly, $p_{x,k} = b - \sum_{k=1}^{i} q_{x,k} = p_{x,i-1} - q_{x,i}$ and $q_{x,i} = b - \sum_{k=1}^{i} p_{x,k} = p_{x,i-1} - p_{x,i}$.

Similar to Lemma 1, the following lemma discusses the relationship between $\pi_y$ and $\theta_y$.

**Lemma 3.** 1. $\pi_{y,i} \leq \pi_{y,j}$ and $\theta_{y,x,j} \leq \theta_{y,i}$ if $j \leq i$.

$$2.\ \pi_{y,i} = b^* - \sum_{k=\beta}^{i} \theta_{y,k} = \pi_{y,i-1} - \theta_{y,i} \tag{11}$$

$$3.\ \theta_{y,i} = b^* - \sum_{k=\beta}^{i} \pi_{y,k} = \pi_{y,i-1} - \pi_{y,i}, \tag{12}$$

**Proof.** Analog to the proofs in Lemmas 1 and 2.

A multi-vector is a vector formed by vectors [26]. To easily represent the feasible solution, the multi-vector $X = (p_x, \pi_y)$ is used. The feasible solution $X$ is a non-reworked feasible solution if $\pi_y$ is a zero vector, e.g., $X = (p_x, 0)$, shown in Fig. 2, for $b = 14$, $p_x = (14, 14, 12, 11, 10)$ and $q_x = (0, 0, 2, 1, 1)$. Otherwise, it is a reworked feasible solution, e.g., $X = (p_x, \pi_y)$, shown in Fig. 3, where $b = 14$, $p_x = (14, 13, 12, 5, 5)$, $q_x = (0, 1, 1, 7, 2)$, $\pi_y = (6, 6, 5, 5)$, and $\theta_y = (0, 0, 1, 0)$.

The following lemma considers a feasible solution is a vector without the rework process.

**Lemma 4.** The solution $X = (p_x, \pi_y)$ is a feasible non-rework solution if $d \leq p_{x,i} \leq p_{x,j}$ for all $j \leq i$ and $\pi_y$ is a zero vector.

**Proof.** Directly follows from Lemma 1 and the definitions of feasible non-rework vector and $X$.

The following lemma considers a feasible solution formed by two vectors with the rework process.

**Lemma 5.** The solution $X = (p_x, \pi_y)$ is a feasible non-rework solution if $p_{x,i} \leq p_{x,j}$ for all $j \leq i$, $\pi_{y,k} \leq \pi_{y,l}$ for all $l \leq k$, $p_{x,n} < d$, $d \leq p_{x,\alpha}$, $\pi_{y,\alpha} \leq q_{x,\alpha} = b - \sum_{k=1}^{\alpha} p_{x,k}$, and $d \leq p_{x,n} + \pi_{y,n}$.

**Proof.** From Lemmas 1 and 3, we have $p_{x,i} \leq p_{x,j}$ for all $j \leq i$, $\pi_{y,a} \leq \pi_{y,b}$ for all $b \leq a$. From the proposed Assumption 1, a work process is necessary if $p_{x,n} < d$. All products outputted from node α are either perfect and sent to next node to process, defective and can be sent back to rework, or must be discarded. Hence, the solution $X$ is required to have at least $d$ units of non-defective products if $d > p_{x,\alpha}$, i.e., we must have $d \leq p_{x,\alpha}$ and also $\pi_{y,\alpha} \leq q_{x,\alpha} = b - \sum_{k=1}^{\alpha} p_{x,k}$ from Lemma 2. A feasible solution must have at least $d$ units of non-defective products from both the first and the second processes in the end, i.e., $d \leq p_{x,n} + \pi_{y,n}$.

## 4. PROPOSED 2-STEP 3-DFS ALGORITHM AND AN EXAMPLE

Based on the innovations, the proposed multi-vector concept and assumptions, the 3-DFS algorithm proposed to calculate the reliability of the proposed general rework network is discussed in this section along with examples to demonstrate the related proposed algorithm.

### 4.1 Proposed 3-DFS concept in find all multi-vectors

To identify all feasible solutions, a 3-DFS is proposed by three depth-first-search methods (DFSs), named $DFS_0$, $DFS_p$, and $DFS_\pi$. $DFS_\theta$. The $DFS_0$ is implemented independently to identify all non-reworked feasible solutions; further, it is similar to the traditional DFS, whose details can be found in [19-21], although with some constraints to reduce the size of solution space; $DFS_p$ and $DFS_\pi$ are implemented in a nested manner to identify the reworked feasible solutions. The proposed 3-DFS is implemented in the following manner:

**STEP D0.** Implement the $DFS_0$ to identify all state vectors; assume $p_x = (b, p_{x,1}, p_{x,2}, \ldots, p_{x,n})$ where $d \leq p_{x,n}$ and $p_{x,i} \leq p_{x,j}$ for all $j \leq I$, and let $I_n$ be the set of such state vectors and $I_r = \emptyset$.

**STEP D1.** Implement the DFS$_p$ to identify a new non-reworked state vector; assume $p_x = (b, p_{x,1}, p_{x,2}, \ldots, p_{x,n})$ for $p_{x,n} < d$ and $p_{x,i} \le p_{x,j}$ for all $j \le i$. If no such vector is identified, then halt.

**STEP D2.** Implement the DFS$_\pi$ to identify a new reworked state vector, assume $\pi_y = (\pi_{y,\alpha}, \pi_{y,\beta}, \pi_{y,\beta+1}, \ldots, \pi_{y,n})$, based on $p_x$, and Lemma 5. If no such vector is identified, then go to STEP D1.

**STEP D3.** Let $I_r = I_r \cup \{(p_x, \pi_y)\}$ and repeat STEP D2.

The traditional DFS is an implicit enumeration, used to identify all possible feasible solutions. From the above, the proposed nested DFS can also identify all feasible solutions. Hence, we have the following lemma.

**Lemma 6.** The proposed 3-DFS can identify all feasible solutions.

**4.2 Occurrent Probability of a Multi-vector**

Any found feasible solution is required to calculate its occurrent probability and the total of all such probability is the required reliability [18, 24]. Let the total number of input products be $b = p_{x,0}$, $p_x = (p_{x,0}, p_{x,1}, p_{x,2}, \ldots, p_{x,n})$, $q_x = (q_{x,0}, q_{x,1}, q_{x,2}, \ldots, q_{x,n})$, $\pi_y = (\pi_{y,\alpha}, \pi_{y,\beta}, \pi_{y,\beta+1}, \ldots, \pi_{y,n})$, and $\theta_y = (\theta_{y,\alpha}, \theta_{y,\beta}, \theta_{y,\beta+1}, \ldots, \theta_{y,n})$. Consider the following two cases:

**Case 1.** The non-rework feasible solution

If $(p_x, \pi_y)$ is a non-rework feasible solution, i.e., $\pi_y$ is a zero vector that does not require the rework process, the number of perfect and defect products in node $i$ are $p_{x,i}$ and $q_{x,i} = b - \sum_{k=1}^{i} p_{x,k} = p_{x,i-1} - p_{x,i}$, respectively, from Lemma 2. Therefore, the related occurrence probability of node $i$ is

$$C_{p_{x,i}}^{p_{x,i-1}}(1-\delta_i)^{p_{x,i}} \delta_i^{q_{x,i}} \cdot C_i(p_{x,i-1}),  \tag{13}$$

where $C_{p_{x,i}}^{p_{x,i-1}} = \dfrac{p_{x,i-1}!}{p_{x,i}! \times q_{x,i}!}$ is the total combination to have $p_{x,i}$ and $q_{x,i}$ units of perfect and defect products from node $i$, $\delta_i$ is the defect rate in node $i$, and $C_i(p_{x,i-1})$ is the probability to have $c_i(p_{x,i-1})$ units of perfect products to input to node $i$ for $i = 1, 2, \ldots, n$. The rework network is constructed in series irrespective of whether it has rework process. Let $p_{x,0} = b$. Thus, the probability of the non-rework feasible solution $(p_x, \pi_y)$ with $\pi_y = 0$ is

$$\Pr((p_x, \pi_y)) = \Pr(p_x) = \prod_{i=1}^{n} C_{p_{x,i}}^{p_{x,i-1}} (1-\delta_i)^{p_{x,i}} \delta_i^{q_{x,i}} \cdot C_i(p_{x,i-1}). \tag{14}$$

**Case 2.** The rework multi-vector

The number of all output perfect products must be larger than or equal to $d$; otherwise, the number of the output reworked perfect-products via $\pi_y$ is required to be $(d - p_{w,n})$ at least.

In the rework process, the number of defects sent from node $\alpha$ to node $\beta$ that required rework is the total number of defects from node $\alpha$ int the first process times a predefined reworked rate, i.e., $b^* = \delta\, q_{y,\alpha}$. Hence, similar to Eq. (14), the occurrent probability of $\pi_y$ is

$$\Pr(\pi_y) = \prod_{i=\beta}^{n} C_{\pi_{y,i}}^{\pi_{y,i-1}} (1-\gamma_i)^{\pi_{y,i}} \gamma_i^{\theta_{y,i}} \cdot C_i(\pi_{y,i-1}), \tag{15}$$

where $\pi_{y,\alpha} = b^*$. The $\pi_y$ is happen right after $p_x$ to complement the number of perfect products and the probability to have $b^*$ reworked perfect products from $q_{y,\alpha}$ units of defect products is

$\Pr(\pi_{y,\alpha}$ units of detective product sent to rework from node $\alpha \mid q_{x,\alpha}$ units of detective product in node $\alpha)$

$$= \Pr(\pi_{y,\alpha} = b^* \mid q_{x,\alpha}) = C_{\pi_{y,\alpha}}^{q_{x,\alpha}} \delta^{\pi_{y,\alpha}} (1-\delta)^{(q_{x,\alpha} - \pi_{y,\alpha})}. \tag{16}$$

Thus,

$$\Pr((p_x, \pi_y)) = \Pr(p_x) \times C_{\pi_{y,\alpha}}^{q_{x,\alpha}} (1-\delta)^{\pi_{y,\alpha}} \delta^{\theta_{y,\alpha}} \times \Pr(\pi_y)$$

$$= \prod_{i=1}^{n} C_{p_{x,i}}^{p_{x,i-1}} (1-\delta_i)^{p_{x,i}} \delta_i^{q_{x,i}} \cdot C_i(p_{x,i-1}) \times C_{\pi_{y,\alpha}}^{q_{x,\alpha}} (1-\delta)^{\pi_{y,\alpha}} \delta^{\theta_{y,\alpha}}$$

$$\times \prod_{i=\beta}^{n} C_{\pi_{y,i}}^{\pi_{y,i-1}} (1-\gamma_i)^{\pi_{y,i}} \gamma_i^{\theta_{y,i}} \cdot C_i(\pi_{y,i-1}). \tag{17}$$

$$= \frac{b!}{q_{x,1}! \cdot q_{x,2}! \cdot \ldots \cdot q_{x,n}!} \prod_{i=1}^{n} (1-\delta_i)^{p_{x,i}} \delta_i^{q_{x,i}} \cdot C_i(p_{x,i-1}) \times C_{\pi_{y,\alpha}}^{q_{x,\alpha}} (1-\delta)^{\pi_{y,\alpha}} \delta^{\theta_{y,\alpha}}$$

$$\times \frac{q_{x,\alpha}!}{\theta_{y,\alpha}!\theta_{y,\beta}!\theta_{y,\beta+1}!\cdots\theta_{y,n}!}\prod_{i=\beta}^{n}(1-\gamma_i)^{\pi_{y,i}}\gamma_i^{\theta_{y,i}}\cdot C_i(\pi_{y,i-1}). \quad (18)$$

From the above, we have the following lemma.

**Lemma 7.** Eq. (16) and Eq. (18) are the occurrent probability of a feasible solution without and with requiring a rework process, respectively.

For example, let $\delta_1=0.05$, $\delta_2=0.10$, $\delta_3=0.15$, $\delta_4=0.20$, $\delta=0.20$, $\gamma_1=0.06$, $\gamma_2=0.10$, $\gamma_3=0.15$, and $\gamma_4=0.20$. In Fig. 2, $X = \Pr((p_x, 0))$ is a feasible solution, where $p_x = (14, 14, 12, 11, 10)$ and we have Table 2:

**Table 2.** Calculation of $\Pr((p_x, 0))$.

| $i$ | $\delta_i$ | $p_{x,i-1}$ | $p_{x,i}$ | $q_{x,i}$ | $C_{p_{x,i}}^{p_{x,i-1}}$ | $(1-\delta_i)^{p_{x,i}}$ | $\delta_i^{q_{x,i}}$ | $c_i(p_{x,i-1})$ | $C_i(p_{x,i-1})$ | $C_{p_{x,i}}^{p_{x,i-1}}(1-\delta_i)^{p_{x,i}}\delta_i^{q_{x,i}}\cdot C_i(p_{x,i-1})$ |
|---|---|---|---|---|---|---|---|---|---|---|
| 1 | 0.05 | 14 | 14 | 0 | 1 | 0.487675 | 1 | 15 | 0.93 | 0.453537731 |
| 2 | 0.10 | 14 | 12 | 2 | 91 | 0.282430 | 0.01 | 15 | 0.95 | 0.244160334 |
| 3 | 0.15 | 12 | 11 | 1 | 12 | 0.167343 | 0.15 | 12 | 0.01 | 0.003012178 |
| 4 | 0.20 | 11 | 10 | 1 | 11 | 0.107374 | 0.2 | 12 | 0.015 | 0.003543348 |

Hence, $\Pr(X) = 0.453537731 \times 0.244160334 \times 0.003012178 \times 0.003543348 = 1.18191\text{E-06}$.

In Fig. 3, $X = (p_x, \pi_y)$ is a feasible solution, where $p_x = (14, 13, 12, 5, 5)$ and $\pi_y = (6, 6, 5, 5)$. Similarly, we have the following results of 1$^{st}$ process and 2$^{nd}$ process in calculating $\Pr(X)$ as summarized in Tables 3 and 4, respectively:

**Table 3.** Calculation of $\Pr(p_x)$.

| $i$ | $\delta_i$ | $p_{x,i-1}$ | $p_{x,i}$ | $q_{x,i}$ | $C_{p_{x,i}}^{p_{x,i-1}}$ | $(1-\delta_i)^{p_{x,i}}$ | $\delta_i^{q_{x,i}}$ | $c_i(p_{x,i-1})$ | $C_i(p_{x,i-1})$ | $C_{p_{x,i}}^{p_{x,i-1}}(1-\delta_i)^{p_{x,i}}\delta_i^{q_{x,i}}\cdot C_i(p_{x,i-1})$ |
|---|---|---|---|---|---|---|---|---|---|---|
| 1 | 0.05 | 14 | 13 | 1 | 14 | 0.513342 | 0.05 | 15 | 0.93 | 0.334185696 |
| 2 | 0.10 | 13 | 12 | 1 | 13 | 0.282430 | 0.10 | 15 | 0.95 | 0.348800478 |
| 3 | 0.15 | 12 | 5 | 7 | 792 | 0.443705 | 1.70859E-06 | 12 | 0.01 | 6.00425E-06 |
| 4 | 0.20 | 5 | 5 | 0 | 1 | 0.327680 | 1.00 | 12 | 0.015 | 0.004915200 |
| | | | | | | | | | Product | 3.44005E-09 |

**Table 4.** Calculation of $\Pr(\pi_y)$.

| $i$ | $\gamma_i$ | $\pi_{y,i-1}$ | $\pi_{y,i}$ | $\theta_{y,i}$ | $C_{\pi_{y,i}}^{\pi_{y,i-1}}$ | $(1-\gamma_i)^{\pi_{y,i}}$ | $\gamma_i^{\theta_{y,i}}$ | $c_i(\pi_{y,i-1})$ | $C_i(\pi_{y,i-1})$ | $C_{\pi_{y,i}}^{\pi_{y,i-1}}(1-\gamma_i)^{\pi_{y,i}}\gamma_i^{\theta_{y,i}} \cdot C_i(\pi_{y,i-1})$ |
|---|---|---|---|---|---|---|---|---|---|---|
| 1* | 0.06 | 6 | 6 | 0 | 1 | 0.689870 | 1.00 | 10 | 0.050 | 0.034493489 |
| 2 | 0.12 | 6 | 5 | 1 | 6 | 0.527732 | 0.12 | 10 | 0.020 | 0.007599340 |
| 3 | 0.18 | 5 | 5 | 0 | 1 | 0.370740 | 1.00 | 6 | 0.010 | 0.003707398 |
| 4 | 0.24 | 5 | 5 | 0 | 1 | 0.253553 | 1.00 | 6 | 0.005 | 0.001267763 |
| | | | | | | | | | Product | 1.23203E-09 |

* Nodes 3 and 2 are nodes α and β in Fig. 3, respectively.

Hence,

$$\Pr(\pi_{y,\alpha}=6 \mid q_{x,\alpha}=7) = C_6^7(1-\delta)^6\delta^1 = 0.531441$$

$$\Pr(X) = \Pr((p_x, \pi_y))$$

$$= \Pr(p_x) \times \Pr(\pi_{y,\alpha}=6 \mid q_{x,\alpha}=7) \times \Pr(\pi_y)$$

$$= 0.334185696 \times 0.348800478 \times 6.00425\text{E-}06 \times 0.004915200 \times 0.531441 \times$$

$$0.034493489 \times 0.007599340 \times 0.003707398 \times 0.001267763$$

$$= 2.25\text{E-}18$$

### 4.3 Proposed Algorithm

The algorithm proposed to estimate the general rework reliability problem is discussed completely in this subsection based on Sections 4.1 and 4.2.

**STEP 1.** Implement STEPs D0–D3 in the proposed 3-DFS discussed in Section 4.1 to estimate all rework multi-vectors.

**STEP 2.** Calculate the occurrent probability of all feasible solutions (events) obtained in STEP 1.

**STEP 3.** Sum up all the occurrent probabilities calculated in SETP 2 and the result is the final reliability required.

From Lemmas 1-7, we have the following theorem:

**Theorem 1.** The proposed algorithm finds all feasible solutions and can calculate the reliability of the general rework network.

## 5. COMPUTSTIONSL EXPERIEMNTS

Before calculating the exact reliability of a general rework network, all feasible solutions must first be obtained [1, 2, 20, 21]. However, all feasible solutions possess a computational difficulty that, in the worst case, raises exponentially with size of the network due to the characteristic of the NP-hard problem [16, 25]. Owing to this inherent problem, a moderate size network shown in Figure 1 was selected to demonstrate this methodology instead of presenting practically large network systems.

Consider the general rework network in Figure 1 of Section 1, where nodes 1, 2, 3, and 4 are the source node, the input node of the rework process (i.e., $\alpha = 3$), first node of the rework process (i.e., $\beta=3$), and sink nodes, respectively. Nodes 0 and 5 are two artificial nodes to represent the input and output number of products in the general rework network. The proposed algorithm is used to obtain all feasible solutions to input $b = 1, 2, …, 15$ unit of products and output at least $d = 1, 2, …, b$ perfect products after processing in the general rework network from nodes 1 to 4. The maximal capacity of each node is 15 and the reliability is zero if $b$ is larger than 15.

To calculate Eqs. (16) and Eq. (18) efficiently, the first step in the procedure is to list all occurrent probability of each state as follows:

**Table 5** Occurrent probability of each state of Example 1.

| State \ Node | 1 | 2 | 3 | 4 |
|---|---|---|---|---|
| 0 | 0.010000 | 0.010000 | 0.005000 | 0.005000 |
| 1 | 0.010000 | 0.020000 | 0.005000 | 0.005000 |
| 2 | 0.010000 | 0.020000 | 0.005000 | 0.005000 |
| 3 | 0.010000 | 0.020000 | 0.005000 | 0.005000 |
| 4 | 0.010000 | 0.020000 | 0.010000 | 0.005000 |
| 5 | 0.010000 | 0.020000 | 0.010000 | 0.005000 |
| 6 | 0.050000 | 0.020000 | 0.010000 | 0.005000 |
| 7 | 0.050000 | 0.020000 | 0.010000 | 0.010000 |
| 8 | 0.050000 | 0.020000 | 0.010000 | 0.010000 |
| 9 | 0.050000 | 0.020000 | 0.010000 | 0.010000 |
| 10 | 0.050000 | 0.020000 | 0.010000 | 0.015000 |
| 11 | 0.930000 | 0.950000 | 0.010000 | 0.015000 |
| 12 | 0.930000 | 0.950000 | 0.010000 | 0.015000 |
| 13 | 0.930000 | 0.950000 | 0.960000 | 0.960000 |
| 14 | 0.930000 | 0.950000 | 0.960000 | 0.960000 |

The second step is to implement the proposed 3-DFS to obtain all feasible solutions, then the summations of these occurrent probabilities based on Eq. (16) or Eq. (18) provides final reliability. The final results for calculating the reliability using the proposed 3-DFS are listed in Tables 6 and 7 for lower and higher defect rate settings: $(\delta_1, \delta_2, \delta_3, \delta_4, \delta, \gamma_1, \gamma_2, \gamma_3, \gamma_4)$ = (0.010, 0.015, 0.020, 0.025, 0.02, 0.012, 0.018, 0.022, 0.029), and (0.05, 0.10, 0.15, 0.20, 0.20, 0.06, 0.12, 0.18, 0.24), respectively.

From Tables 6 and 7, we draw the following observations:

1. Irrespective of the values of $\delta_i$, $\delta$, and $\gamma_i$ for $i$ = 1, 2, 3, 4, the numbers of $N_n$ and $N_r$ both are always constants, e.g., $N_n$ = 35 for $(b, d)$ = (4, 1) and $N_r$ = 35 for $(b, d)$ = (5, 2) in Tables 6 and 7.

2. Because a lower defect rate has a higher chance for lesser number of defects than that of the higher defect rate, the setting with the lower defect rate has better $R_n$, $R_r$, and $R$ than those of the setting with a higher defect rate. The above observation is confirmed from the ratio of $R_h/R_l$, which is always larger than 1, listed in the last column in Table 7.

3. The value of $R_h/R_l$ is increasing with the increment of values $b$ and $d$, respectively. Surprisingly, the value of $R_h/R_l$ is up to 1181.42449 for $b = d = 15$.

4. The larger $b$ the higher $N_n$, $N_r$, $R_n$, $R_r$, $R_l$ and $R_h$ for fixed $d$ owing to the factor that the occurrent probability is increasing with the value of state, e.g., the occurrent probabilities of states 10 and 11 are 0.05 and 0.93 of node 1, respectively.

5. The values of $N_n$, $R_n$, $R_l$ and $R_h$ are decreased with the increasement of $d$, e.g., $N_n$ = 15, 5, 1, $R_n$ = 0.00000000499841136288, 0.00000000493331982160, 0.00000000404460769381 if $d$ = 1, 2, 3 for $b$ = 3, respectively, in Tables 6 and 7. The above observation is owing to the larger d the lower chance to achieve the same.

6. The values of $N_r$ and $R_r$ are increased up to a peak then decreased with the increasement of $d$, e.g., $N_r$ = 84, 112, 105, 80, 50, 24, 7, if $d$ = 1, 2, 3, 4, 5, 6, 7 for $b$ = 7, respectively, in Tables 6 and 7.

7. The larger $N_n$ and $N_r$, the higher values of $R_n$ and $R_r$, respectively, and vice versa.

8. The smaller $N_n$ and $N_r$, the lower values of $R_n$ and $R_r$, respectively, and vice versa.

9. The larger $b$ (but must not larger than the maximal capacity of each node), the greater number of $N_n$ and $N_r$, respectively.

10. If $b$ is larger than the maximal capacity of any node, both values of $R_n$ and $R_r$ are zero.

11. Interestingly, at least 99.999979% and 99.999799% of $R_l$ and $R_h$ are obtained from $R_n$ from Tables 6 and 7, respectively, i.e., it appears as if the rework process is not very useful in increasing the network reliability under these assumptions provided in this paper.

Table 6. Final results obtained from the low defect rate setting of Figure 1.

| $b$ | $d$ | $N_n$ | $N_r$ | $R_n$ | $R_r$ | $R_l = R_n + R_r$ | $R_n/R_l$ | $R_r/R_l$ |
|---|---|---|---|---|---|---|---|---|
| 1 | 1 | 1 | 1 | 4.6588E-09 | 5.4590E-17 | 4.6588E-09 | 0.99999999 | 1.1718E-08 |
| 2 | 1 | 5 | 4 | 4.9767E-09 | 7.4500E-18 | 4.9767E-09 | 1.00000000 | 1.4970E-09 |
|   | 2 | 1 | 2 | 4.3408E-09 | 1.0291E-16 | 4.3408E-09 | 0.99999998 | 2.3707E-08 |
| 3 | 1 | 15 | 10 | 4.9984E-09 | 7.6000E-19 | 4.9984E-09 | 1.00000000 | 1.5205E-10 |
|   | 2 | 5 | 8 | 4.9333E-09 | 2.1040E-17 | 4.9333E-09 | 1.00000000 | 4.2649E-09 |
|   | 3 | 1 | 3 | 4.0446E-09 | 1.4553E-16 | 4.0446E-09 | 0.99999996 | 3.5981E-08 |
| 4 | 1 | 35 | 20 | 9.5211E-09 | 9.0000E-20 | 9.5211E-09 | 1.00000000 | 9.4527E-12 |
|   | 2 | 15 | 20 | 9.5137E-09 | 4.0300E-18 | 9.5137E-09 | 1.00000000 | 4.2360E-10 |
|   | 3 | 5 | 12 | 9.3433E-09 | 6.4790E-17 | 9.3433E-09 | 0.99999999 | 6.9344E-09 |
|   | 4 | 1 | 4 | 7.5372E-09 | 3.6585E-16 | 7.5372E-09 | 0.99999995 | 4.8539E-08 |
| 5 | 1 | 70 | 35 | 9.9706E-09 | 1.0000E-20 | 9.9706E-09 | 1.00000000 | 1.0029E-12 |
|   | 2 | 35 | 40 | 9.9698E-09 | 5.8000E-19 | 9.9698E-09 | 1.00000000 | 5.8175E-11 |
|   | 3 | 15 | 30 | 9.9464E-09 | 1.2620E-17 | 9.9464E-09 | 1.00000000 | 1.2688E-09 |
|   | 4 | 5 | 16 | 9.5946E-09 | 1.2439E-16 | 9.5946E-09 | 0.99999999 | 1.2965E-08 |
|   | 5 | 1 | 5 | 7.0228E-09 | 4.3116E-16 | 7.0228E-09 | 0.99999994 | 6.1394E-08 |
| 6 | 1 | 126 | 56 | 4.9993E-08 | 0.0000E+00 | 4.9993E-08 | 1.00000000 | 0.0000E+00 |
|   | 2 | 70 | 70 | 4.9992E-08 | 3.4000E-19 | 4.9992E-08 | 1.00000000 | 6.8010E-12 |
|   | 3 | 35 | 60 | 4.9979E-08 | 9.3500E-18 | 4.9979E-08 | 1.00000000 | 1.8708E-10 |
|   | 4 | 15 | 40 | 4.9728E-08 | 1.2719E-16 | 4.9728E-08 | 1.00000000 | 2.5577E-09 |
|   | 5 | 5 | 20 | 4.7096E-08 | 8.7853E-16 | 4.7096E-08 | 0.99999998 | 1.8654E-08 |
|   | 6 | 1 | 6 | 3.2718E-08 | 2.4391E-15 | 3.2718E-08 | 0.99999993 | 7.4551E-08 |
| 7 | 1 | 210 | 84 | 8.6396E-08 | 0.0000E+00 | 8.6396E-08 | 1.00000000 | 0.0000E+00 |
|   | 2 | 126 | 112 | 8.6396E-08 | 4.0000E-20 | 8.6396E-08 | 1.00000000 | 4.6299E-13 |
|   | 3 | 70 | 105 | 8.6394E-08 | 1.2500E-18 | 8.6394E-08 | 1.00000000 | 1.4469E-11 |
|   | 4 | 35 | 80 | 8.6363E-08 | 2.2630E-17 | 8.6363E-08 | 1.00000000 | 2.6203E-10 |
|   | 5 | 15 | 50 | 8.5926E-08 | 2.3538E-16 | 8.5926E-08 | 1.00000000 | 2.7393E-09 |
|   | 6 | 5 | 24 | 8.2071E-08 | 1.5594E-15 | 8.2071E-08 | 0.99999998 | 1.9001E-08 |
|   | 7 | 1 | 7 | 6.0970E-08 | 5.3664E-15 | 6.0970E-08 | 0.99999991 | 8.8018E-08 |
| 8 | 1 | 330 | 120 | 9.7696E-08 | 0.0000E+00 | 9.7696E-08 | 1.00000000 | 0.0000E+00 |
|   | 2 | 210 | 168 | 9.7696E-08 | 0.0000E+00 | 9.7696E-08 | 1.00000000 | 0.0000E+00 |
|   | 3 | 126 | 168 | 9.7696E-08 | 1.6000E-19 | 9.7696E-08 | 1.00000000 | 1.6377E-12 |
|   | 4 | 70 | 140 | 9.7692E-08 | 3.7300E-18 | 9.7692E-08 | 1.00000000 | 3.8181E-11 |
|   | 5 | 35 | 100 | 9.7628E-08 | 5.1730E-17 | 9.7628E-08 | 1.00000000 | 5.2987E-10 |
|   | 6 | 15 | 60 | 9.6827E-08 | 4.9693E-16 | 9.6827E-08 | 0.99999999 | 5.1322E-09 |
|   | 7 | 5 | 28 | 9.0096E-08 | 2.9094E-15 | 9.0096E-08 | 0.99999997 | 3.2292E-08 |

| | | | | | | | |
|---|---|---|---|---|---|---|---|
| | 8 | 1 | 8 | 5.6809E-08 | 5.7833E-15 | 5.6809E-08 | 0.99999990 | 1.0180E-07 |
| 9 | 1 | 495 | 165 | 9.9701E-08 | 0.0000E+00 | 9.9701E-08 | 1.00000000 | 0.0000E+00 |
| | 2 | 330 | 240 | 9.9701E-08 | 0.0000E+00 | 9.9701E-08 | 1.00000000 | 0.0000E+00 |
| | 3 | 210 | 252 | 9.9701E-08 | 1.0000E-20 | 9.9701E-08 | 1.00000000 | 1.0030E-13 |
| | 4 | 126 | 224 | 9.9700E-08 | 5.6000E-19 | 9.9700E-08 | 1.00000000 | 5.6168E-12 |
| | 5 | 70 | 175 | 9.9691E-08 | 9.7100E-18 | 9.9691E-08 | 1.00000000 | 9.7400E-11 |
| | 6 | 35 | 120 | 9.9554E-08 | 1.1604E-16 | 9.9554E-08 | 1.00000000 | 1.1656E-09 |
| | 7 | 15 | 70 | 9.8046E-08 | 8.9156E-16 | 9.8046E-08 | 0.99999999 | 9.0932E-09 |
| | 8 | 5 | 32 | 8.7824E-08 | 3.5235E-15 | 8.7824E-08 | 0.99999996 | 4.0120E-08 |
| | 9 | 1 | 9 | 5.2932E-08 | 6.1355E-15 | 5.2932E-08 | 0.99999988 | 1.1591E-07 |
| 10 | 1 | 715 | 220 | 1.3173E-07 | 0.0000E+00 | 1.3173E-07 | 1.00000000 | 0.0000E+00 |
| | 2 | 495 | 330 | 1.3173E-07 | 0.0000E+00 | 1.3173E-07 | 1.00000000 | 0.0000E+00 |
| | 3 | 330 | 360 | 1.3173E-07 | 0.0000E+00 | 1.3173E-07 | 1.00000000 | 0.0000E+00 |
| | 4 | 210 | 336 | 1.3173E-07 | 7.0000E-20 | 1.3173E-07 | 1.00000000 | 5.3138E-13 |
| | 5 | 126 | 280 | 1.3173E-07 | 1.5900E-18 | 1.3173E-07 | 1.00000000 | 1.2070E-11 |
| | 6 | 70 | 210 | 1.3171E-07 | 2.2780E-17 | 1.3171E-07 | 1.00000000 | 1.7296E-10 |
| | 7 | 35 | 140 | 1.3144E-07 | 2.1804E-16 | 1.3144E-07 | 1.00000000 | 1.6589E-09 |
| | 8 | 15 | 80 | 1.2906E-07 | 1.2735E-15 | 1.2906E-07 | 0.99999999 | 9.8675E-09 |
| | 9 | 5 | 36 | 1.1643E-07 | 4.8630E-15 | 1.1643E-07 | 0.99999996 | 4.1769E-08 |
| | 10 | 1 | 10 | 7.3980E-08 | 9.6439E-15 | 7.3980E-08 | 0.99999987 | 1.3036E-07 |
| 11 | 1 | 1001 | 286 | 1.1679E-04 | 0.0000E+00 | 1.1679E-04 | 1.00000000 | 0.0000E+00 |
| | 2 | 715 | 440 | 1.1679E-04 | 0.0000E+00 | 1.1679E-04 | 1.00000000 | 0.0000E+00 |
| | 3 | 495 | 495 | 1.1679E-04 | 0.0000E+00 | 1.1679E-04 | 1.00000000 | 0.0000E+00 |
| | 4 | 330 | 480 | 1.1679E-04 | 2.0100E-18 | 1.1679E-04 | 1.00000000 | 1.7211E-14 |
| | 5 | 210 | 420 | 1.1679E-04 | 9.5000E-17 | 1.1679E-04 | 1.00000000 | 8.1345E-13 |
| | 6 | 126 | 336 | 1.1679E-04 | 1.8210E-15 | 1.1679E-04 | 1.00000000 | 1.5593E-11 |
| | 7 | 70 | 245 | 1.1677E-04 | 2.4261E-14 | 1.1677E-04 | 1.00000000 | 2.0777E-10 |
| | 8 | 35 | 160 | 1.1654E-04 | 2.1163E-13 | 1.1654E-04 | 1.00000000 | 1.8159E-09 |
| | 9 | 15 | 90 | 1.1458E-04 | 1.3312E-12 | 1.1458E-04 | 0.99999999 | 1.1618E-08 |
| | 10 | 5 | 40 | 1.0293E-04 | 5.4140E-12 | 1.0293E-04 | 0.99999995 | 5.2600E-08 |
| | 11 | 1 | 11 | 6.0901E-05 | 8.8396E-12 | 6.0901E-05 | 0.99999985 | 1.4515E-07 |
| 12 | 1 | 1365 | 364 | 1.3118E-04 | 0.0000E+00 | 1.3118E-04 | 1.00000000 | 0.0000E+00 |
| | 2 | 1001 | 572 | 1.3118E-04 | 0.0000E+00 | 1.3118E-04 | 1.00000000 | 0.0000E+00 |
| | 3 | 715 | 660 | 1.3118E-04 | 0.0000E+00 | 1.3118E-04 | 1.00000000 | 0.0000E+00 |
| | 4 | 495 | 660 | 1.3118E-04 | 0.0000E+00 | 1.3118E-04 | 1.00000000 | 0.0000E+00 |
| | 5 | 330 | 600 | 1.3118E-04 | 2.0000E-17 | 1.3118E-04 | 1.00000000 | 1.5246E-13 |
| | 6 | 210 | 504 | 1.3118E-04 | 4.6000E-16 | 1.3118E-04 | 1.00000000 | 3.5066E-12 |
| | 7 | 126 | 392 | 1.3118E-04 | 6.8800E-15 | 1.3118E-04 | 1.00000000 | 5.2448E-11 |
| | 8 | 70 | 280 | 1.3112E-04 | 6.8856E-14 | 1.3112E-04 | 1.00000000 | 5.2515E-10 |
| | 9 | 35 | 180 | 1.3052E-04 | 5.0382E-13 | 1.3052E-04 | 1.00000000 | 3.8601E-09 |
| | 10 | 15 | 100 | 1.2629E-04 | 2.5291E-12 | 1.2629E-04 | 0.99999998 | 2.0026E-08 |
| | 11 | 5 | 44 | 1.0662E-04 | 7.1586E-12 | 1.0662E-04 | 0.99999993 | 6.7142E-08 |
| | 12 | 1 | 12 | 5.6744E-05 | 9.0956E-12 | 5.6744E-05 | 0.99999984 | 1.6029E-07 |
| 13 | 1 | 1820 | 455 | 4.5361E-01 | 0.0000E+00 | 4.5361E-01 | 1.00000000 | 0.0000E+00 |
| | 2 | 1365 | 728 | 4.5361E-01 | 0.0000E+00 | 4.5361E-01 | 1.00000000 | 0.0000E+00 |
| | 3 | 1001 | 858 | 4.5361E-01 | 0.0000E+00 | 4.5361E-01 | 1.00000000 | 0.0000E+00 |
| | 4 | 715 | 880 | 4.5361E-01 | 0.0000E+00 | 4.5361E-01 | 1.00000000 | 0.0000E+00 |
| | 5 | 495 | 825 | 4.5361E-01 | 0.0000E+00 | 4.5361E-01 | 1.00000000 | 0.0000E+00 |
| | 6 | 330 | 720 | 4.5361E-01 | 0.0000E+00 | 4.5361E-01 | 1.00000000 | 0.0000E+00 |
| | 7 | 210 | 588 | 4.5361E-01 | 5.5012E-14 | 4.5361E-01 | 1.00000000 | 1.2127E-13 |
| | 8 | 126 | 448 | 4.5361E-01 | 1.1130E-12 | 4.5361E-01 | 1.00000000 | 2.4536E-12 |
|   | 9  | 70   | 315  | 4.5361E-01| 2.4161E-11| 4.5361E-01| 1.00000000 | 5.3264E-11|
|   | 10 | 35   | 200  | 4.5349E-01| 3.8042E-10| 4.5349E-01| 1.00000000 | 8.3886E-10|
|   | 11 | 15   | 110  | 4.5180E-01| 2.4999E-09| 4.5180E-01| 0.99999999 | 5.5333E-09|
|   | 12 | 5    | 48   | 4.3453E-01| 1.6981E-08| 4.3453E-01| 0.99999996 | 3.9080E-08|
|   | 13 | 1    | 13   | 3.2485E-01| 5.7107E-08| 3.2485E-01| 0.99999982 | 1.7580E-07|
| 14| 1  | 2380 | 560  | 7.1278E-01| 0.0000E+00| 7.1278E-01| 1.00000000 | 0.0000E+00|
|   | 2  | 1820 | 910  | 7.1278E-01| 0.0000E+00| 7.1278E-01| 1.00000000 | 0.0000E+00|
|   | 3  | 1365 | 1092 | 7.1278E-01| 0.0000E+00| 7.1278E-01| 1.00000000 | 0.0000E+00|
|   | 4  | 1001 | 1144 | 7.1278E-01| 0.0000E+00| 7.1278E-01| 1.00000000 | 0.0000E+00|
|   | 5  | 715  | 1100 | 7.1278E-01| 0.0000E+00| 7.1278E-01| 1.00000000 | 0.0000E+00|
|   | 6  | 495  | 990  | 7.1278E-01| 0.0000E+00| 7.1278E-01| 1.00000000 | 0.0000E+00|
|   | 7  | 330  | 840  | 7.1278E-01| 2.9976E-14| 7.1278E-01| 1.00000000 | 4.2055E-14|
|   | 8  | 210  | 672  | 7.1278E-01| 7.2597E-13| 7.1278E-01| 1.00000000 | 1.0185E-12|
|   | 9  | 126  | 504  | 7.1278E-01| 1.6065E-11| 7.1278E-01| 1.00000000 | 2.2539E-11|
|   | 10 | 70   | 350  | 7.1270E-01| 2.6518E-10| 7.1270E-01| 1.00000000 | 3.7207E-10|
|   | 11 | 35   | 220  | 7.1154E-01| 2.0163E-09| 7.1154E-01| 1.00000000 | 2.8336E-09|
|   | 12 | 15   | 120  | 6.9906E-01| 1.3408E-08| 6.9906E-01| 0.99999998 | 1.9180E-08|
|   | 13 | 5    | 52   | 6.1304E-01| 5.3848E-08| 6.1304E-01| 0.99999991 | 8.7838E-08|
|   | 14 | 1    | 14   | 3.0268E-01| 5.8016E-08| 3.0268E-01| 0.99999981 | 1.9168E-07|
| 15| 1  | 3060 | 680  | 7.9285E-01| 0.0000E+00| 7.9285E-01| 1.00000000 | 0.0000E+00|
|   | 2  | 2380 | 1120 | 7.9285E-01| 0.0000E+00| 7.9285E-01| 1.00000000 | 0.0000E+00|
|   | 3  | 1820 | 1365 | 7.9285E-01| 0.0000E+00| 7.9285E-01| 1.00000000 | 0.0000E+00|
|   | 4  | 1365 | 1456 | 7.9285E-01| 0.0000E+00| 7.9285E-01| 1.00000000 | 0.0000E+00|
|   | 5  | 1001 | 1430 | 7.9285E-01| 0.0000E+00| 7.9285E-01| 1.00000000 | 0.0000E+00|
|   | 6  | 715  | 1320 | 7.9285E-01| 0.0000E+00| 7.9285E-01| 1.00000000 | 0.0000E+00|
|   | 7  | 495  | 1155 | 7.9285E-01| 7.1054E-15| 7.9285E-01| 1.00000000 | 8.9619E-15|
|   | 8  | 330  | 960  | 7.9285E-01| 2.5402E-13| 7.9285E-01| 1.00000000 | 3.2039E-13|
|   | 9  | 210  | 756  | 7.9285E-01| 5.8811E-12| 7.9285E-01| 1.00000000 | 7.4177E-12|
|   | 10 | 126  | 560  | 7.9282E-01| 1.0050E-10| 7.9282E-01| 1.00000000 | 1.2676E-10|
|   | 11 | 70   | 385  | 7.9239E-01| 8.7117E-10| 7.9239E-01| 1.00000000 | 1.0994E-09|
|   | 12 | 35   | 240  | 7.8747E-01| 6.0260E-09| 7.8747E-01| 0.99999999 | 7.6524E-09|
|   | 13 | 15   | 130  | 7.5071E-01| 2.7440E-08| 7.5071E-01| 0.99999996 | 3.6552E-08|
|   | 14 | 5    | 56   | 5.9186E-01| 5.8556E-08| 5.9186E-01| 0.99999990 | 9.8936E-08|
|   | 15 | 1    | 15   | 2.8202E-01| 5.8643E-08| 2.8202E-01| 0.99999979 | 2.0794E-07|

**Table 7.** The final results obtained from the higher defect rate setting of Figure 1.

| $b$ | $d$ | $N_n$ | $N_r$ | $R_n$ | $R_r$ | $R_h = R_n + R_r$ | $R_n/R_h$ | $R_r/R_h$ | $R_h/R_l$ |
|---|---|---|---|---|---|---|---|---|---|
| 1 | 1 | 1  | 1  | 2.9070E-09 | 1.3347E-16 | 2.9070E-09 | 0.99999995 | 4.5913E-08 | 1.602607 |
| 2 | 1 | 5  | 4  | 4.1239E-09 | 1.9555E-16 | 4.1239E-09 | 0.99999997 | 4.7419E-08 | 1.206807 |
| 2 | 2 | 1  | 2  | 1.6901E-09 | 8.3800E-17 | 1.6901E-09 | 0.99999988 | 4.9582E-08 | 2.568350 |
| 3 | 1 | 15 | 10 | 4.6333E-09 | 2.4492E-16 | 4.6333E-09 | 0.99999998 | 5.2861E-08 | 1.078813 |
| 3 | 2 | 5  | 8  | 3.1051E-09 | 1.8005E-16 | 3.1051E-09 | 0.99999992 | 5.7985E-08 | 1.588776 |
| 3 | 3 | 1  | 3  | 9.8264E-10 | 6.1200E-17 | 9.8264E-10 | 0.99999982 | 6.2281E-08 | 4.116056 |
| 4 | 1 | 35 | 20 | 7.4904E-09 | 2.9253E-16 | 7.4904E-09 | 0.99999999 | 3.9054E-08 | 1.271098 |
| 4 | 2 | 15 | 20 | 6.3994E-09 | 4.9571E-16 | 6.3994E-09 | 0.99999995 | 7.7462E-08 | 1.486650 |
| 4 | 3 | 5  | 12 | 3.8634E-09 | 2.9507E-16 | 3.8634E-09 | 0.99999987 | 7.6377E-08 | 2.418445 |
| 4 | 4 | 1  | 4  | 1.1426E-09 | 4.0750E-17 | 1.1426E-09 | 0.99999974 | 3.5664E-08 | 6.596421 |
| 5 | 1 | 70 | 35 | 9.1295E-09 | 2.5059E-16 | 9.1295E-09 | 1.00000000 | 2.7448E-08 | 1.092133 |
| 5 | 2 | 35 | 40 | 8.4291E-09 | 5.8979E-16 | 8.4291E-09 | 0.99999997 | 6.9971E-08 | 1.182788 |
| 5 | 3 | 15 | 30 | 6.2929E-09 | 6.1375E-16 | 6.2929E-09 | 0.99999991 | 9.7531E-08 | 1.580577 |
| 5 | 4 | 5  | 16 | 3.0558E-09 | 2.2699E-16 | 3.0558E-09 | 0.99999980 | 7.4281E-08 | 3.139804 |
| 5 | 5 | 1 | 5 | 6.6432E-10 | 1.2084E-16 | 6.6432E-10 | 0.99999966 | 1.8190E-07 | 10.571472 |
| 6 | 1 | 126 | 56 | 4.8695E-08 | 8.8507E-16 | 4.8695E-08 | 1.00000000 | 1.8176E-08 | 1.026648 |
| 6 | 2 | 70 | 70 | 4.6711E-08 | 2.6108E-15 | 4.6711E-08 | 0.99999998 | 5.5892E-08 | 1.070256 |
| 6 | 3 | 35 | 60 | 3.9406E-08 | 3.8510E-15 | 3.9406E-08 | 0.99999993 | 9.7727E-08 | 1.268331 |
| 6 | 4 | 15 | 40 | 2.5290E-08 | 2.8296E-15 | 2.5290E-08 | 0.99999985 | 1.1189E-07 | 1.966337 |
| 6 | 5 | 5 | 20 | 1.0274E-08 | 8.3956E-16 | 1.0274E-08 | 0.99999972 | 8.1720E-08 | 4.584119 |
| 6 | 6 | 1 | 6 | 1.9312E-09 | 6.6500E-17 | 1.9312E-09 | 0.99999957 | 3.4435E-08 | 16.941917 |
| 7 | 1 | 210 | 84 | 5.4984E-08 | 5.6785E-16 | 5.4984E-08 | 1.00000000 | 1.0327E-08 | 1.571278 |
| 7 | 2 | 126 | 112 | 5.3950E-08 | 2.0456E-15 | 5.3950E-08 | 0.99999999 | 3.7916E-08 | 1.601409 |
| 7 | 3 | 70 | 105 | 4.9484E-08 | 4.0243E-15 | 4.9484E-08 | 0.99999996 | 8.1326E-08 | 1.745917 |
| 7 | 4 | 35 | 80 | 3.8846E-08 | 4.6527E-15 | 3.8846E-08 | 0.99999990 | 1.1977E-07 | 2.223213 |
| 7 | 5 | 15 | 50 | 2.3565E-08 | 3.2738E-15 | 2.3565E-08 | 0.99999980 | 1.3893E-07 | 3.646289 |
| 7 | 6 | 5 | 24 | 9.8691E-09 | 1.2097E-15 | 9.8691E-09 | 0.99999967 | 1.2257E-07 | 8.315923 |
| 7 | 7 | 1 | 7 | 2.2456E-09 | 3.5150E-17 | 2.2456E-09 | 0.99999946 | 1.5653E-08 | 27.151237 |
| 8 | 1 | 330 | 120 | 6.5488E-08 | 3.4823E-16 | 6.5488E-08 | 1.00000000 | 5.3174E-09 | 1.491803 |
| 8 | 2 | 210 | 168 | 6.4974E-08 | 1.5170E-15 | 6.4974E-08 | 0.99999999 | 2.3348E-08 | 1.503612 |
| 8 | 3 | 126 | 168 | 6.2400E-08 | 3.8212E-15 | 6.2400E-08 | 0.99999998 | 6.1237E-08 | 1.565628 |
| 8 | 4 | 70 | 140 | 5.4974E-08 | 6.0594E-15 | 5.4974E-08 | 0.99999993 | 1.1022E-07 | 1.777059 |
| 8 | 5 | 35 | 100 | 4.1174E-08 | 6.4095E-15 | 4.1174E-08 | 0.99999985 | 1.5567E-07 | 2.371085 |
| 8 | 6 | 15 | 60 | 2.3738E-08 | 4.0151E-15 | 2.3738E-08 | 0.99999973 | 1.6914E-07 | 4.078963 |
| 8 | 7 | 5 | 28 | 8.8255E-09 | 8.5538E-16 | 8.8255E-09 | 0.99999955 | 9.6921E-08 | 10.208548 |
| 8 | 8 | 1 | 8 | 1.3056E-09 | 1.8230E-17 | 1.3056E-09 | 0.99999934 | 1.3963E-08 | 43.512766 |
| 9 | 1 | 495 | 165 | 7.6755E-08 | 2.0880E-16 | 7.6755E-08 | 1.00000000 | 2.7204E-09 | 1.298950 |
| | 2 | 330 | 240 | 7.6505E-08 | 1.0846E-15 | 7.6505E-08 | 1.00000000 | 1.4177E-08 | 1.303189 |
| | 3 | 210 | 252 | 7.5067E-08 | 3.3490E-15 | 7.5067E-08 | 0.99999999 | 4.4614E-08 | 1.328156 |
| | 4 | 126 | 224 | 7.0122E-08 | 6.6440E-15 | 7.0122E-08 | 0.99999995 | 9.4750E-08 | 1.421817 |
| | 5 | 70 | 175 | 5.8733E-08 | 8.9587E-15 | 5.8734E-08 | 0.99999989 | 1.5253E-07 | 1.697353 |
| | 6 | 35 | 120 | 4.0331E-08 | 7.5576E-15 | 4.0331E-08 | 0.99999978 | 1.8739E-07 | 2.468436 |
| | 7 | 15 | 70 | 1.9843E-08 | 3.1922E-15 | 1.9843E-08 | 0.99999962 | 1.6087E-07 | 4.941111 |
| | 8 | 5 | 32 | 5.6777E-09 | 5.9657E-16 | 5.6777E-09 | 0.99999944 | 1.0507E-07 | 15.468339 |
| | 9 | 1 | 9 | 7.5906E-10 | 9.3700E-18 | 7.5906E-10 | 0.99999921 | 1.2344E-08 | 69.733869 |
| 10 | 1 | 715 | 220 | 8.8007E-08 | 1.2324E-16 | 8.8007E-08 | 1.00000000 | 1.4003E-09 | 1.496841 |
| | 2 | 495 | 330 | 8.7887E-08 | 7.4775E-16 | 8.7887E-08 | 1.00000000 | 8.5081E-09 | 1.498886 |
| | 3 | 330 | 360 | 8.7097E-08 | 2.7217E-15 | 8.7097E-08 | 0.99999999 | 3.1249E-08 | 1.512477 |
| | 4 | 210 | 336 | 8.3922E-08 | 6.4031E-15 | 8.3922E-08 | 0.99999997 | 7.6299E-08 | 1.569700 |
| | 5 | 126 | 280 | 7.5214E-08 | 1.0333E-14 | 7.5214E-08 | 0.99999991 | 1.3738E-07 | 1.751423 |
| | 6 | 70 | 210 | 5.8309E-08 | 1.0972E-14 | 5.8309E-08 | 0.99999982 | 1.8818E-07 | 2.258821 |
| | 7 | 35 | 140 | 3.5485E-08 | 7.0890E-15 | 3.5485E-08 | 0.99999969 | 1.9977E-07 | 3.703966 |
| | 8 | 15 | 80 | 1.5306E-08 | 2.8938E-15 | 1.5306E-08 | 0.99999954 | 1.8906E-07 | 8.431914 |
| | 9 | 5 | 36 | 4.3910E-09 | 6.1768E-16 | 4.3910E-09 | 0.99999934 | 1.4067E-07 | 26.514404 |
| | 10 | 1 | 10 | 6.6197E-10 | 1.4266E-15 | 6.6198E-10 | 0.99999907 | 2.1551E-06 | 111.755993 |
| 11 | 1 | 1001 | 286 | 5.4447E-05 | 2.4166E-14 | 5.4447E-05 | 1.00000000 | 4.4384E-10 | 2.144970 |
| | 2 | 715 | 440 | 5.4431E-05 | 1.8803E-13 | 5.4431E-05 | 1.00000000 | 3.4546E-09 | 2.145596 |
| | 3 | 495 | 495 | 5.4296E-05 | 8.6854E-13 | 5.4296E-05 | 1.00000000 | 1.5996E-08 | 2.150932 |
| | 4 | 330 | 480 | 5.3590E-05 | 2.5921E-12 | 5.3590E-05 | 0.99999998 | 4.8369E-08 | 2.179266 |
| | 5 | 210 | 420 | 5.1084E-05 | 5.3501E-12 | 5.1084E-05 | 0.99999995 | 1.0473E-07 | 2.286153 |
| | 6 | 126 | 336 | 4.4844E-05 | 7.6519E-12 | 4.4844E-05 | 0.99999988 | 1.7063E-07 | 2.604232 |
| | 7 | 70 | 245 | 3.3924E-05 | 7.5005E-12 | 3.3924E-05 | 0.99999977 | 2.2109E-07 | 3.441984 |
| | 8 | 35 | 160 | 2.0622E-05 | 5.1480E-12 | 2.0622E-05 | 0.99999964 | 2.4964E-07 | 5.651206 |
| | 9 | 15 | 90 | 9.3522E-06 | 2.1930E-12 | 9.3522E-06 | 0.99999945 | 2.3449E-07 | 12.251617 |
| | 10 | 5 | 40 | 2.7182E-06 | 3.7369E-13 | 2.7182E-06 | 0.99999919 | 1.3748E-07 | 37.866728 |
|    | 11 | 1    | 11   | 3.4003E-07 | 1.4482E-15 | 3.4003E-07 | 0.99999890 | 4.2590E-09 | 179.100943 |
| 12 | 1  | 1365 | 364  | 8.9657E-05 | 2.6125E-14 | 8.9657E-05 | 1.00000000 | 2.9139E-10 | 1.463165   |
|    | 2  | 1001 | 572  | 8.9641E-05 | 2.1477E-13 | 8.9641E-05 | 1.00000000 | 2.3959E-09 | 1.463423   |
|    | 3  | 715  | 660  | 8.9499E-05 | 1.0530E-12 | 8.9499E-05 | 1.00000000 | 1.1766E-08 | 1.465735   |
|    | 4  | 495  | 660  | 8.8717E-05 | 3.3729E-12 | 8.8717E-05 | 0.99999999 | 3.8019E-08 | 1.478656   |
|    | 5  | 330  | 600  | 8.5772E-05 | 7.5774E-12 | 8.5772E-05 | 0.99999996 | 8.8344E-08 | 1.529431   |
|    | 6  | 210  | 504  | 7.7936E-05 | 1.2079E-11 | 7.7936E-05 | 0.99999990 | 1.5499E-07 | 1.683202   |
|    | 7  | 126  | 392  | 6.3054E-05 | 1.3661E-11 | 6.3054E-05 | 0.99999981 | 2.1666E-07 | 2.080412   |
|    | 8  | 70   | 280  | 4.2791E-05 | 1.1205E-11 | 4.2791E-05 | 0.99999968 | 2.6184E-07 | 3.064113   |
|    | 9  | 35   | 180  | 2.2807E-05 | 6.1485E-12 | 2.2807E-05 | 0.99999951 | 2.6959E-07 | 5.722868   |
|    | 10 | 15   | 100  | 8.5751E-06 | 1.8568E-12 | 8.5751E-06 | 0.99999928 | 2.1653E-07 | 14.727093  |
|    | 11 | 5    | 44   | 1.9058E-06 | 2.5432E-13 | 1.9058E-06 | 0.99999903 | 1.3345E-07 | 55.945032  |
|    | 12 | 1    | 12   | 1.9770E-07 | 8.6499E-14 | 1.9770E-07 | 0.99999871 | 4.3753E-07 | 287.028429 |
| 13 | 1  | 1820 | 455  | 1.4329E-02 | 2.2470E-13 | 1.4329E-02 | 1.00000000 | 1.5681E-11 | 31.656033  |
|    | 2  | 1365 | 728  | 1.4329E-02 | 1.4669E-12 | 1.4329E-02 | 1.00000000 | 1.0237E-10 | 31.656079  |
|    | 3  | 1001 | 858  | 1.4329E-02 | 5.7525E-12 | 1.4329E-02 | 1.00000000 | 4.0145E-10 | 31.656625  |
|    | 4  | 715  | 880  | 1.4327E-02 | 2.8621E-11 | 1.4327E-02 | 1.00000000 | 1.9977E-09 | 31.660863  |
|    | 5  | 495  | 825  | 1.4316E-02 | 1.2093E-10 | 1.4316E-02 | 1.00000000 | 8.4470E-09 | 31.684772  |
|    | 6  | 330  | 720  | 1.4269E-02 | 4.1427E-10 | 1.4269E-02 | 0.99999999 | 2.9033E-08 | 31.789874  |
|    | 7  | 210  | 588  | 1.4101E-02 | 1.1507E-09 | 1.4101E-02 | 0.99999997 | 8.1605E-08 | 32.169235  |
|    | 8  | 126  | 448  | 1.3605E-02 | 2.4971E-09 | 1.3605E-02 | 0.99999992 | 1.8355E-07 | 33.342711  |
|    | 9  | 70   | 315  | 1.2399E-02 | 3.8569E-09 | 1.2399E-02 | 0.99999980 | 3.1108E-07 | 36.585199  |
|    | 10 | 35   | 200  | 1.0060E-02 | 3.9747E-09 | 1.0060E-02 | 0.99999962 | 3.9511E-07 | 45.079821  |
|    | 11 | 15   | 110  | 6.6155E-03 | 2.8966E-09 | 6.6155E-03 | 0.99999940 | 4.3785E-07 | 68.293593  |
|    | 12 | 5    | 48   | 3.0333E-03 | 1.0582E-09 | 3.0333E-03 | 0.99999905 | 3.4885E-07 | 143.250067 |
|    | 13 | 1    | 13   | 7.0619E-04 | 3.9920E-13 | 7.0620E-04 | 0.99999850 | 5.6528E-10 | 459.993765 |
| 14 | 1  | 2380 | 560  | 6.2264E-02 | 1.5070E-12 | 6.2264E-02 | 1.00000000 | 2.4203E-11 | 11.447697  |
|    | 2  | 1820 | 910  | 6.2264E-02 | 5.5998E-12 | 6.2264E-02 | 1.00000000 | 8.9936E-11 | 11.447703  |
|    | 3  | 1365 | 1092 | 6.2264E-02 | 1.8300E-11 | 6.2264E-02 | 1.00000000 | 2.9391E-10 | 11.447783  |
|    | 4  | 1001 | 1144 | 6.2260E-02 | 8.6079E-11 | 6.2260E-02 | 1.00000000 | 1.3826E-09 | 11.448502  |
|    | 5  | 715  | 1100 | 6.2235E-02 | 3.9494E-10 | 6.2235E-02 | 1.00000000 | 6.3460E-09 | 11.453166  |
|    | 6  | 495  | 990  | 6.2107E-02 | 1.4774E-09 | 6.2107E-02 | 0.99999999 | 2.3788E-08 | 11.476660  |
|    | 7  | 330  | 840  | 6.1592E-02 | 4.4128E-09 | 6.1592E-02 | 0.99999998 | 7.1646E-08 | 11.572662  |
|    | 8  | 210  | 672  | 5.9896E-02 | 1.0134E-08 | 5.9896E-02 | 0.99999993 | 1.6920E-07 | 11.900320  |
|    | 9  | 126  | 504  | 5.5406E-02 | 1.6517E-08 | 5.5406E-02 | 0.99999982 | 2.9810E-07 | 12.864665  |
|    | 10 | 70   | 350  | 4.6108E-02 | 1.8129E-08 | 4.6108E-02 | 0.99999964 | 3.9318E-07 | 15.457368  |
|    | 11 | 35   | 220  | 3.1653E-02 | 1.4067E-08 | 3.1653E-02 | 0.99999943 | 4.4440E-07 | 22.479387  |
|    | 12 | 15   | 120  | 1.5757E-02 | 6.1265E-09 | 1.5757E-02 | 0.99999911 | 3.8880E-07 | 44.363767  |
|    | 13 | 5    | 52   | 4.5492E-03 | 7.1386E-10 | 4.5492E-03 | 0.99999865 | 1.5692E-07 | 134.758684 |
|    | 14 | 1    | 14   | 4.1058E-04 | 9.8702E-13 | 4.1058E-04 | 0.99999826 | 2.4039E-09 | 737.189202 |
| 15 | 1  | 3060 | 680  | 1.5129E-01 | 5.1850E-12 | 1.5129E-01 | 1.00000000 | 3.4273E-11 | 5.240724   |
|    | 2  | 2380 | 1120 | 1.5129E-01 | 1.6350E-11 | 1.5129E-01 | 1.00000000 | 1.0807E-10 | 5.240726   |
|    | 3  | 1820 | 1365 | 1.5129E-01 | 4.2000E-11 | 1.5129E-01 | 1.00000000 | 2.7762E-10 | 5.240745   |
|    | 4  | 1365 | 1456 | 1.5128E-01 | 1.6718E-10 | 1.5128E-01 | 1.00000000 | 1.1051E-09 | 5.240936   |
|    | 5  | 1001 | 1430 | 1.5124E-01 | 7.6874E-10 | 1.5124E-01 | 1.00000000 | 5.0829E-09 | 5.242292   |
|    | 6  | 715  | 1320 | 1.5103E-01 | 3.0206E-09 | 1.5103E-01 | 0.99999999 | 2.0001E-08 | 5.249771   |
|    | 7  | 495  | 1155 | 1.5008E-01 | 9.4572E-09 | 1.5008E-01 | 0.99999998 | 6.3016E-08 | 5.282971   |
|    | 8  | 330  | 960  | 1.4672E-01 | 2.2639E-08 | 1.4672E-01 | 0.99999994 | 1.5430E-07 | 5.403878   |
|    | 9  | 210  | 756  | 1.3730E-01 | 3.8655E-08 | 1.3730E-01 | 0.99999984 | 2.8153E-07 | 5.774505   |
|    | 10 | 126  | 560  | 1.1686E-01 | 4.5134E-08 | 1.1686E-01 | 0.99999967 | 3.8621E-07 | 6.784198   |
|    | 11 | 70   | 385  | 8.3667E-02 | 3.7545E-08 | 8.3667E-02 | 0.99999946 | 4.4874E-07 | 9.470721   |

| 12 | 35 | 240 | 4.5222E-02 | 1.9164E-08 | 4.5222E-02 | 0.99999917 | 4.2378E-07 | 17.413409 |
| 13 | 15 | 130 | 1.5810E-02 | 4.4246E-09 | 1.5810E-02 | 0.99999879 | 2.7987E-07 | 47.483897 |
| 14 | 5 | 56 | 2.8168E-03 | 1.5943E-16 | 2.8168E-03 | 0.99999843 | 5.6601E-14 | 210.120915 |
| 15 | 1 | 15 | 2.3871E-04 | 1.3347E-16 | 2.3871E-04 | 0.99999799 | 5.5912E-13 | 1181.42449 |

## 6. CONCLUSIONS

Network reliability is a popular method for validating the designs and evaluating the performance of several practical systems. More than 25 previous studies have analyzed the rework network reliability problem [1, 2]. However, these studies are either based on faulty assumptions as mentioned in [1, 2] and Section 2.4, or errors are observed in their calculations as discussed in Sections 2.1–2.3. Moreover, these assumptions and limitations in the traditional rework network are not reasonable and limit the applications of the rework network. Hence, a new general rework network is proposed making better assumptions without too many limitations along with a new algorithm based on three DFSs and a few simple properties of the problem to solve its reliability.

The experiments discussed in Section 5 show that the proposed algorithm not only helps us calculate the reliability of the general rework network, but it also aids in understanding the relationships among the values of $b$, $d$, $N_n$, $N_r$, $R_n$, $R_r$, $R$, $\delta_i$, $\delta$, and $\gamma_i$ for all $i$. Hence, the proposed algorithm helps decision makers make better decisions regarding the rework process and can be extended for assessing other stochastic reliability problems.